\documentstyle[prl,twocolumn,aps]{revtex}
\def\be{\begin{equation}}
\def\ee{\end{equation}}
\def\bea{\begin{eqnarray}}
\def\eea{\end{eqnarray}}
\def\bma{\begin{mathletters}}
\def\ema{\end{mathletters}}
\def\C{\hbox{$\mit I$\kern-.7em$\mit C$}}
\def\one{\hbox{$\mit I$\kern-.6em$\mit I$}}
\font\Bbb =msbm10  scaled \magstephalf
\def\id{{\hbox{\Bbb I}}}

\begin{document}
\draft

\title{Bound entanglement for continuous variables is
a rare phenomenon}

\author{Pawe\l\ Horodecki$^1$, J. Ignacio Cirac$^2$,  and  Maciej
Lewenstein$^3$}

\address{$^1$ Faculty of Applied Physics and Mathematics\\
Technical University of Gda\'nsk, 80--952 Gda\'nsk, Poland}
\address{$^2$ Institut f\"ur Theoretische Physik, Universit\"at Innsbruck,
A-6020 Innsbruck, Austria}
\address{$^3$ Institut
f\"ur Theoretische Physik, Universit\"at Hannover\\
 30167 Hannover, Germany}

\maketitle

\begin{abstract}
We discuss the notion of bound entanglement (BE) for continuous
variables (CV). We show that the set of non--distillable states
(NDS) for CV is nowhere dense in the set of all states, i.e.,
the states of infinite--dimensional bipartite systems are
generically distillable. This automatically implies that the
sets of separable states, entangled states with positive partial
transpose, and bound entangled states are also nowhere dense in
the set of all states. All these properties significantly
distinguish quantum CV systems from the spin like ones.
The aspects of the definition of BE for CV
is also analysed, especially in context of
Schmidt numbers theory. In particular the main result
is generalised by means of arbitrary Schmidt
number and single copy regime.
\end{abstract}

\narrowtext

%%%%%%%%%%%%%%%%%%%%%%%%
\section {Introduction}
%%%%%%%%%%%%%%%%%%%%%%%%

Bound entanglement \cite{bound} is the entanglement which cannot
be distilled, i.e. no pure state entanglement can be obtained
from it by means of local operations and classical communication
(LOCC)\cite{dist}. So far, it has been studied mainly for spin
like systems. These studies has allowed to discover many
interesting properties of bound entanglement, for both
bipartite\cite{aktyw}, and multiparticle systems\cite{multi}.
Recently, much attention has been devoted continuous variable
(CV) systems (c.f. \cite{Bra}). Bound entanglement has also been
considered for continuous variables (CV), and the first
nontrivial examples of BES for CV have been
constructed\cite{Ho00} (see also \cite{Werner}). Once we have
some examples of BES for CV, it is interesting to ask how
frequent is the phenomenon of bound entanglement, i. e. how many
states of that kind are in set of all CV states?

The question of "how many quantum states having some interesting
property are there?" is very natural. In the context of
entanglement it was first considered in Ref. \cite{volume},
where the problem of the volume of the subset of separable
(non--entangled) states in the set of all bipartite states of
spin systems was considered. Numerical evidence has shown that
the volume of the set of separable states approaches zero when
the size of the spin goes to infinity. It was also shown that
for any finite spin system the volume of separable states is
nonzero due to the existence of a separable neighborhood, i.e.
an open ball of separable states in the vicinity of the
maximally mixed state in arbitrary dimension.
%@
Further first analystical bounds on the size of
neighbourhood have been provided \cite{TV}.
All this raised a series of questions concerning the interpretation
of experiments of quantum computing based on high temperature NMR; many
interesting analyses have been performed in this context
\cite{Caves,Popescu}.

The question of the ``size'' of the set representing separable
states has been recently answered \cite{CH00} for CV; it has
been show that for bipartite states this subset is nowhere dense
(relative to the trace--norm topology). This implies that this
set does not contain any open ball and also that CV states are
generically non--separable. On the other hand there exists
another subset that is of interest in the context of
entanglement. This is the subset of of non--distillable states
(NDS), i.e. states that cannot be distilled. This subset
contains the separable states, and therefore it might well be
that an appreciable fraction of all states are in such a subset.
In this paper we show that this is not the case; that is, the
subset of NDS is nowhere dense in the set of bipartite states of
CV. We present two different proofs of this fact. One uses the
uniform topology, and the other one the trace--norm topology.

We also perform analsysis how one can relax conditions
of NDS in context of CV in comparison with the standard definition,
and prove stronger version of the main result with help of Schmidt numbers
theory \cite{Te00} and single copy regime (see \cite{single,xor}).

There are several results which follow from our proofs. In
particular, since the subset of NDS contains the subset of BES,
we have that that subset is also nowhere dense. The same thing
occurs with the subset of states with positive partial transpose
(PPT) \cite{bound} and therefore with with those PPT states that
are entangled. Moreover, since the subset of separable states is
also contained in the one of NDS, our results include the ones
given in Ref.\ \cite{CH00}.

%%%%%%%%%%%%%%%%%%%%%%%%%%%%%%%%%%%%%%%%%%%%%%%%%%%%%%%%%%%%%%%%%%%%%%%%
\section{Bound entangled states for continuous variables}
%%%%%%%%%%%%%%%%%%%%%%%%%%%%%%%%%%%%%%%%%%%%%%%%%%%%%%%%%%%%%%%%%%%%%%%%

The main subject of this paper is the question about whether
generic CV states are non--distillable. We present below the
answer to this question: the subset of NDS is nowhere dense. In
this section we first discuss the definition of NDS. Then, we
present two proofs of our result. First, it will be proven using
the uniform topology and exploiting the fact that any density
operator can be considered as the limit of a sequence of density
operators defined on a finite support. Then, following the
approach of Ref. \cite{CH00} we shall prove the general
statement that any proper closed subset of bipartite states
which is invariant under local transformations is nowhere dense
(in the trace norm topology). This generalizes the results of
Ref. \cite{CH00} and, as we shall see, together with the fact
that the set of NDS is closed, proves our claims.

\subsection{Bound entangled states}

Let us denote by $M$ set of all density operators acting on
$H_A\otimes H_B$, where $H_A\cong H_B \cong L^2(R)$. Let us
consider a density operator $\rho\in M$. According to Ref.\
\cite{bound}, $\rho$ is distillable \cite{bound} iff there
exists some finite $n\in N$, two rank two projectors $P_A,P_B$
acting on $H_A,H_B$, respectively, such that
\be
\label{ddd}
((P_A\otimes P_B)\rho^{\otimes n}(P_A\otimes
P_B))^{T_B}
\not\ge 0;
\ee
otherwise, $\rho$ is non--distillable. We call $D$ and $N$ the
set of all distillable and non--distillable density operators,
respectively. Physically, this definition tells us that a state
is distillable iff out of a sufficiently large number of copies
we can obtain by local operations two qubits which are
entangled. The reason for that is clear. First, given the fact
that one can distill maximally entangled states out of all
entangled states of qubits, this means that if the above
condition is true, we can always distill maximally entangled
qubit states out of the original state $\rho$. Second, if the
above condition is not fulfilled for any $n$, then we will not
be able to produce (asymptotically) any qubit maximally
entangled state by using local operations alone.

In the following we will reexpress Eq. (\ref{ddd}) as follows:
\be
\label{ddd2}
\epsilon\equiv \langle \Psi|((P_A\otimes P_B)\rho^{\otimes n}(P_A\otimes
P_B))^{T_B}|\Psi\rangle < 0
\ee
for some $|\Psi\rangle\in H_A\otimes H_B$. Without loosing
generality we can take $|\Psi\rangle=P_A\otimes
P_B^{T_B}|\Psi\rangle$, i.e. $|\Psi\rangle$ belongs to the
$2\otimes2$ subspace determined by $P_A, P_B^{T_B}$. It is easy
to see then that
\be
\label{decomp}
(|\Psi\rangle\langle\Psi|)^{T_B}=\sum_{k=1}^4\lambda_k
|\phi_k\rangle\langle\phi_k|,
\ee
where $-1/2\le\lambda_k\le 1$, i.e. $|\lambda_k|\le 1$ and the
$|\phi_k\rangle$ are also in the same $2\times 2$ subspace.

%%%%%%%%%%%%%%%%%%%%%%%%%%%%%%%%%%%%%%%%%%%%%%%%%%%%%%%%%%%%%%%%%%%%%%%
\subsection{Bound entangled states for CV  are nowhere dense: Proof I}
%%%%%%%%%%%%%%%%%%%%%%%%%%%%%%%%%%%%%%%%%%%%%%%%%%%%%%%%%%%%%%%%%%%%%%%

In this subsection we show that the set $N$ of NDS states is
{\em nowhere dense} in the set of all states $M$. We will first
recall some definitions. A subset $A$ of a topological space $X$
is nowhere dense if its closure contains no open set. Note that
if $A$ is already closed then it is nowhere dense iff it
contains no open set. For example, the subset of integer numbers
$Z$ in $R$ (with the topology induced by the absolute value
metric) is nowhere dense since it is already close and no
neighborhood of any integer contains only integers.

In this subsection we will use the uniform topology for
operators, which is the one derived from the operator norm.
First we will show that $N$ is closed with that topology by
proving that its complement, $D$, is open. Then, we will show
that $D$ is dense in the set of all density operators. From this
last point it follows that $N$ (equivalently, its closure)
contains no open set and therefore it is nowhere dense.

In order to show that $D$ is open, let us consider some $\rho\in
D$. According to the definition of $D$, there exists some finite
integer $n$, and a Schmidt rank two state $|\Psi\rangle\in
H_A\otimes H_B$ such that Eq. (\ref{ddd2}) is fulfilled with
(\ref{decomp}). Let us consider an open ball
$B_\eta(\rho)=\{\rho', ||\rho'-\rho||<\eta\}$. We will show that
for $\eta < |\epsilon|/4n$, $B_\eta(\rho)\subset D$.

To this aim we argue that
\bea
&&\left|\langle \Psi|((P_A\otimes P_B)(\rho^{\otimes
n}-(\rho')^{\otimes n})(P_A\otimes P_B))^{T_B}|\Psi\rangle
\right |\nonumber\\
 &&=\left|\sum_{k=1}^4\lambda_k\langle\phi_k|
\rho^{\otimes n}-(\rho')^{\otimes n}|\phi_k\rangle\right|\nonumber\\
 &&\le 4||\rho^{\otimes n}-(\rho')^{\otimes n}||\le 4\eta n.
\eea
The latter inequality can be proven by induction, using the identity
\bea
\rho^{\otimes n}-(\rho')^{\otimes
n}&=&\frac{1}{2}(\rho^{\otimes n-1}+(\rho')^{\otimes
n-1})\otimes(\rho-\rho') \nonumber\\
&+& \frac{1}{2}(\rho^{\otimes
n-1}-(\rho')^{\otimes n-1})\otimes(\rho+\rho') ,
\eea
and the fact that both $||\rho||$ and $||\rho'||$ are smaller
than one. Thus, if $\eta<|\epsilon|/4n$, we see that for any
$\rho'\in B_\eta(\rho)$,
\be
\langle \Psi|((P_A\otimes P_B)(\rho')^{\otimes n}(P_A\otimes
P_B))^{T_B}|\Psi\rangle<0,
\ee
{\it ergo} $\rho'$ is distillable.

Now we show that $D$ is dense in $M$. To this aim we observe
that for any $\rho\in M$ we can always find a sequence
$\{\rho_n\}_{n=0}^\infty$, with $\rho_n\in D$ such that $\rho_n
\stackrel{u}{\rightarrow} \rho$. We consider the spectral
decomposition of $\rho$ as
\be
\rho = \sum_{n=1}^\infty p_n |\Psi_n\rangle\langle \Psi_n|,
\ee
where we have chosen $p_1\ge p_2,\ldots$. Note that since $\rho$
is a trace class operator, the sequence $p_n$ converges
monotonically to zero. On the other hand, we can write the
Schmidt decomposition of each $|\Psi_n\rangle$ as
\be
|\Psi_n\rangle = \sum_{k=1}^\infty \sqrt{\lambda_{n,k}}
|u_{n,k},v_{n,k}\rangle,
\ee
where again we have chosen $\lambda_{n,k}\ge \lambda_{n,k+1}\ge
0$ and $\lambda_{n,k}$ converges monotonically to zero as $k\to
\infty$. Now, we define
\be
\tilde \rho_N \equiv \sum_{n=1}^N p_n
|\Psi_{N,n}\rangle\langle \Psi_{N,n}|,
\ee
where
\be
|\Psi_{N,n}\rangle = \sum_{k=1}^N \sqrt{\lambda_{n,k}}
|u_{n,k},v_{n,k}\rangle.
\ee
It is clear that $\tilde\rho_N$ is supported on $H_A^{N}\otimes
H_B^{N}$, where both $H_A^{N}$ have finite dimension.
Thus, we can always find two pairs of orthogonal vectors
$|a_{1,2}^{N}\rangle\in H_A\ominus H_A^{N}$ and
$|b_{1,2}^{N}\rangle\in H_B\ominus H_B^{N}$. Let us define
$|\Phi^{N}\rangle = (|a_1,b_1\rangle+|a_2,b_2\rangle)/\sqrt{2}$
and
\be
\rho_N = K_N (\tilde \rho_N + \frac{1}{N} |\Phi^{N}\rangle\langle \Phi^{N}|),
\ee
where $K_N$ is a normalization constant. It is clear that
$\rho_N\stackrel{N\to\infty}{\rightarrow} \rho$. On the other
hand, defining
\bma
\bea
P_A^{N} &=& |a_1\rangle\langle a_1| + |a_2\rangle\langle a_2|,\\
P_B^{N} &=& |b_1\rangle\langle b_1| + |b_2\rangle\langle b_2|
\eea
\ema
and taking $n=1$ it is clear that $\rho_N\in D$.
This completes the proof. $\Box$

Obviously the fact that the set of NDS states for CV is nowhere
dense, implies that the contained in it sets of BES and PPT
states are also nowhere dense. One can, however, prove the
latter directly using the method used above. The only difference
would be that the state $|\Psi\rangle$ which in above proof
belongs to a $2\otimes 2 $ subspace has to be substituted by a
general vector $|\Psi\rangle$ of arbitrary Schmidt rank, but at
the same time there is no need to consider $n$-fold tensor
products, since the PPT property of $\rho$ is maintained for
arbitrary number of its copies. One has to use, however, the
property of the partially transposed projector
$||(|\Psi\rangle\langle\Psi|)^{T_B}||\le 1$.

\subsection{Bound entangled states for CV  are nowhere dense: Proof II}

In order to demonstrate the statement of the previous section in
another way, we first need to recall notions of several
necessary tools, which have been used in Ref. \cite{CH00}. Let
${\cal B} ({\cal H}_{1})\otimes {\cal B}({\cal H}_{2}) $ stand
for bounded operators on Hilbert space ${\cal H} ={\cal
H}_{1}\otimes {\cal H}_{2}$ describing our bipartite system. We
assume that at least one of the subsystems is described by CV,
and hence it has the infinite dimension.

Now, consider a third auxiliary system described by ${\cal
H}_{3}$. It is convenient to describe all states in ${\cal
H}={\cal H}_{1}\otimes {\cal H}_{2}$ as reduced states of some
{\it pure} states in the extended space ${\cal H} \otimes {\cal
H}_{3}$. If we have a pure state $|v_{123}\rangle \langle
v_{123}|$ in the extended space, then the reduced state ${\rm
Tr}_3(|v_{123}\rangle \langle v_{123}|)$ is denoted by
$\varrho_{12}$. Let us denote by ${\cal T}$ the set of all
states on ${\cal B} ({\cal H}_{1})\otimes {\cal B}({\cal H}_{2})
$. This is a set of unit trace operators with nonnegative
spectrum. We shall also endow this set with the norm topology
$|| \cdot ||_{T}$ , $||A||_{T}\equiv {\rm
Tr}(\sqrt{A^{\dagger}A})$. Now, one defines \cite{CH00} the map
$\Phi$ from the unit sphere ${\cal S}$ representing all
wavefunctions form ${\cal H} \otimes {\cal H}_{3}$ to the set of
states ${\cal T}$ in the following way :
\begin{equation}
\Phi(v_{123})={\rm Tr}_{3}(| v_{123}\rangle \langle v_{123}|)=
\varrho_{12}.
\end{equation}
The map $\Phi: {\cal S} \rightarrow {\cal T}$ is {\it continuous
(in the norm $|| \cdot||_T$) and onto}. In particular it maps
dense subsets onto dense subsets (see \cite{CH00} for
explanation).

Consider the set ${\cal X}$ of all vectors $u_{123}=A \otimes I
\otimes I (v_{123})$ for all $A \in {\cal B} ({\cal H}_{1})$.
The vector $v_{123}$ is called {\it 1-cyclic } (see \cite{CH00})
if the closure of ${\cal X} $ in the norm $|| \cdot ||_{T}$
turns out to be {\it the whole} space ${\cal H}_{1} \otimes
{\cal H}_{2} \otimes {\cal H}_{3}$. The physical interpretation
of 1-cyclic vectors in both finite dimensional, as well as in
the CV case, is that those are the vectors which have maximal
possible Schmidt rank. Note, that according to Lemma 2 of Ref.
[25] they form a dense set in ${\cal H}_1\otimes{\cal
H}_2\otimes{\cal H}_3$.

Now we consider the following simple

{\it Observation 1.- Let the set ${\cal ND}$ be (i) a proper
closed (in $|| \cdot ||_{T}$ norm) subset of the set of states
${\cal T}$ which is (ii) invariant under the operations $A
\otimes I$. Then any vector $v_{123}$ satisfying
$\Phi(v_{123})\in {\cal ND}$ cannot be 1-cyclic
%!
and ${\cal ND}$ is nowhere dense in ${\cal T}$.}
%%!

The above observation is a natural generalization of the Lemma 1
of Ref. \cite{CH00}. To show this, consider such vector $v$ that
its ``reduction'' $\Phi(v)$ belongs to ${\cal ND}$, and take any
vector
\begin{equation}
v'=A \otimes I \otimes I (v),
\label{v'}
\end{equation}
defined for arbitrary $A$, such that $||v'||=1$. We shall show
first that $\Phi(v')$ also belongs to ${\cal ND}$. Indeed (see
\cite{CH00}) we have $\Phi(v')=A \otimes I \Phi(v) A^{\dagger}
\otimes I$ and (because the norm of $v'$ is one) the trace
$\Phi(v')$ is one. But, because the set ${\cal ND}$ is closed
under the operation $A \otimes I ( \cdot) A^{\dagger} \otimes
I$, we see that $\Phi(v')$ still belongs to the set.

Now suppose that $v$ were 1-cyclic. Then, that the set ${\cal
M}$ of all vectors $v'$ would be dense in the unit sphere ${\cal
S}$ of all normalized vectors belonging to ${\cal H} \otimes
{\cal H}_{3}$. As the map $\Phi$ is continuous and onto, it
certainly would map ${\cal M}$ onto some new set denoted by
$\Phi({\cal M})$, which would be dense in set of all bipartite
states ${\cal T}$. Thus closure of $\Phi({\cal M})$ must have
give all ${\cal T}$. But, on the other hand any element of
$\Phi({\cal M})$ (which is defined as $\Phi(v')$ for some vector
$v'$ of the form (\ref{v'})) belongs to ${\cal ND}$. As the
latter is closed, the closure of $\Phi({\cal M})$ would have to
be a subset of ${\cal ND}$. But ${\cal ND}$ was supposed to be
closed and strictly smaller than the set ${\cal T}$, so the
closure of $\Phi({\cal M})$ cannot be equal to ${\cal T}$. This
gives the required contradiction. The above reasoning follows
the lines of the proof of Ref. \cite{CH00}. The only difference
is that instead of the specific set of separable states
considered there, here we have considered an abstract set ${\cal
ND}$, which has some special properties. Note, that the
assumptions of Observation 1 and the fact that the 1-cyclic vectors form a
dense set in ${\cal S}$ imply that the closed set ${\cal ND}$ is nowhere
dense. If it had contained a open set,
%!
then,
following continuity of $\Phi$
this open set would have
had to be an image of
open  subset of ${\cal H}_1\otimes{\cal
H}_2\otimes{\cal H}_3$, which would have had to contain a ball,
an thus a 1-cyclic vector.
%The latter follows from the fact that
%1-cyclic vectors form a dense set in ${\cal S}$.
%!!
Now, to show that the set of NDS states is nowhere dense we
have to show that it is (i) invariant under local operations of
the type $A\otimes I$, (ii) closed in the trace norm $|| \cdot
||_T$. The first property (i) is immediate, since a NDS cannot
be converted into a free entangled state by means of local
operations. The second one is not so obvious for continuous
variables, but it follows from the results of the previous
subsection. We thus have:

{\it Observation 2.- The property of non--distillability is
invariant under the one side local action $A \otimes I (\cdot )
A^{\dagger}\otimes I$.}

The proof is simple - the arguments of Ref. \cite{bound} can be
applied (see also \cite{Ho00}) to show that any local separable
superoperator cannot cause that the state looses the
non--distillability property.

{\it Observation 3.- The set of all NDS is closed in the norm
$||\cdot ||_{T}$ .}

To prove the closeness of the set of NDS, we prove that its
complement, i.e. the set of distillable states $D$ is open in
the trace norm. To this aim we repeat the arguments of
subsection A and consider some $\rho\in D$, for which there
exists some finite integer number $n$, $P_A,P_B$, rank two
projectors acting on $H_A,H_B$, and a rank two vector
$|\Psi\rangle\in H_A\otimes H_B$ such that Eq. (\ref{ddd2}) is
fulfilled. We consider now an open ball in the trace norm, i.e.
$\tilde{B}_\eta(\rho)=\{\rho', ||\rho'-\rho||_T<\eta\}$. Note that if
$\rho'\in \tilde{B}_\eta(\rho)$ then the operator norm fulfills
$||\rho'-\rho||\le ||\rho'-\rho||_T<\eta$ \cite{oper}. Using the
same argument as before we show that for $\eta < |\epsilon|/4n$,
$\tilde{B}_\eta(\rho)\subset D$, which completes the proof. $\Box$

Combining the Observations 1.-3. we see that the set of NDS
states is nowhere dense in the trace norm, which implies the
same property for the BES, PPT states, and separable states.
%!
%%%%%%%%%%%%%%%%%%%%%
\section{Analysis}
%%%%%%%%%%%%%%%%%%%%%
%!!

As mentioned in the introduction, non--trivial BES for
continuous variables (CV) have been discovered. In this
subsection we discuss some of the details regarding these
states, as well as whether entangled states in CV with infinite
Schmidt number represent are generic in the set of entangled
states.

\subsection{Continuous variable bound entangled states}

The construction of non-trivial BES for CV systems \cite{Ho00}
was based on an idea similar to the one used for spin systems, for which it
has been proven that any entangled state with positive partial
transpose\cite{Asher} cannot be distilled\cite{bound}. The
crucial element of the construction was to create the state in
such a way that it cannot be obtained simply by embedding a
bound entangled state in a finite dimensional Hilbert space into
the Hilbert space of CV.

The particular example $\varrho$ of CV BES proposed by us was
first of all assumed to satisfy the condition that its partial
transpose $\varrho^{T_{B}}$, defined as
\begin{equation}
\varrho^{T_B}_{m\mu,n\nu}\equiv
\langle m, \mu| \varrho^{T_B}| n,\nu \rangle=
\varrho_{m\nu,n\mu},
\label{transposition}
\end{equation}
has a nonnegative spectrum. Such requirement was, however, not
sufficient, as one could invent the following "trivial" example
of a PPT entangled states for CV\cite{Ho00}
\begin{equation}
\tilde{\sigma}=\mathop{\oplus}\limits_{n=1}^{\infty}p_n \sigma_n.
\label{oszuk}
\end{equation}
The above state is build from infinitely many ``copies'' of the
same $3 \otimes 3$ \footnote{Subsequently we shall denote by $n
\otimes n$ states the states of quantum systems defined on the
Hilbert space ${\cal H}={\cal C}^{n} \otimes {\cal C}^{n}$. The
space will be sometimes called ``$n \otimes n$ space''.} BES
$\sigma$ labeled by $\sigma_n$. Each of $\sigma_{n}$ has the
matrix elements of the original $\sigma$, but in the basis
$S_n=\{ |i,j\rangle \}_{i,j=3n}^{3n+3} $. Here $\{ p_i
\}_{i=1}^{\infty}$ is an infinite sequence of nonzero
probabilities, $\sum_{i=1}^{\infty} p_i=1$. The bound
entanglement of the CV state $\tilde\sigma$ is in a certain
sense spurious, as it can in principle be reversibly converted
by means of local operations and classical communication into
the $3 \otimes 3$ entanglement.

One could easily construct another example, similar to the one
above, with $\sigma_{n}$ acting in $k_{n} \otimes k_{n}$ Hilbert
space with $k_{n} \rightarrow \infty$, and $\tilde\sigma$ being
block diagonal as in (\ref{oszuk}).
This example is much more interesting as far as CV are concerned,
because it can not be reversibly converted into any state
of fixed spin. Thus to some extend it might be regarded as
generic BE. However, such a state would
still be a mixture of ``locally orthogonal'' spin states, which
does not exploit fully the CV Hilbert space structure, i.e.
infinite dimension fully. In fact, if such CV BES were produced
by a random mixture, they could be easily ``decoupled '' by
local projective measurements and classical communication.

%@
Thus we propose to define generic BES for CV
in a stronger way, namely as the states from
which no pure entanglement can be distilled, and they
cannot be represented by the states of the above sort.
%@

This is a somewhat phenomenological definition, but it implies to single
out some required properties of the generic CV BES. The first
nontrivial examples of the generic CV BES, presented in Ref.
\cite{Ho00},
%
%were in fact constructed in order to fulfill those
%
fulfill those requirements.
These states have the form:
\begin{equation}
\varrho\propto |\Psi \rangle \langle \Psi | +
\sum_{n=1}^{\infty}
\sum_{m>n}^{\infty} |\Psi_{mn} \rangle \langle \Psi_{mn}|,
\label{stan}
\end{equation}
with the following definitions of the symbols: $|\Psi \rangle=
\sum_{n=1}^{\infty} a_n |n,n \rangle$,
$|\Psi\rangle \in {\cal H} = l^2({\cal C}) \otimes l^2({\cal
C})$ with the finite norm $||\Psi||^2=\sum_{n=1}^{\infty}
|a_n|^2=q< \infty$, and vectors
\begin{equation}
|\Psi_{mn} \rangle= c_m a_n |n,m \rangle +(c_m)^{-1} a_m |m,n
\rangle,
\label{cztery}
\end{equation}
for $n<m$ with (in general) complex $a_n$ and $c_n$, such that
(i) $0<|c_{n+1}|<|c_n|<1$, (ii) $\sum_{n=1}^{\infty}
\sum_{m>n}^{\infty} ||\Psi_{mn} ||^2 $ is finite.
The latter condition can be achieved for example by setting
$a_n=a^{n}$, $c_n=c^{n}$, for some $0<a<c<1$, see \cite{Ho00}.
Physically, the vector $|\Psi\rangle$, when normalized, may
describe a state of two modes of the quantized electromagnetic
field, or more generally two harmonic oscillators. The state
(\ref{stan}) has the following properties: (i) it is bound
entangled, as it has the PPT property (i. e. it has the positive
partial transpose); (ii) it is not a simple ``direct sum'' of
finite spin BES in a sense of the ``spurious'' examples
discussed above (Eq. (\ref{oszuk})).

Recently considerable attention has been devoted to the so
called Gaussian states. In systems of two harmonic oscillator
modes (one of Alice, one of Bob), i.e. in the, so called,
Gaussian $1\times 1$ case, it has been shown that no bound
entanglement exists -- such Gaussian states are either separable
\cite{Duan00,Si00}, or distillable \cite{Gi00}. In another
words, in this case PPT property is a necessary and sufficient
condition for separability, and non-distillability. This result
can be extended to the case $1\times N$. Soon after realizing
this facts Werner and Wolf have found an example of a Gaussian
BES with PPT property \cite{Werner}. This result has been
achieved by considering first covariance matrices of Gaussian
states and their null subspaces. It was noted that the Gaussian
state is separable, iff its covariance matrix can be minorized
by some block diagonal covariance matrix. Second, the
characterization of PPT states in terms of covariance matrix has
been found. The BES has been constructed using an elegant
explicit construction, performed using the analysis of the range
and the ``subtraction method'' first developed for spin systems
in Refs. \cite{Le98,substrac,iha1,iha2}. In the terminology of
Refs. \cite{Le98,substrac,iha1,iha2} the states found in Ref.
\cite{Werner} are examples of the so called "edge states". The
approach of Ref. \cite{Werner} can be used further to analyze
multiparticle entanglement. In particular, one can try to
``split'' the covariance matrix of $n \times n$ state in a way
to get $m \times m \times m $ state with some bound entanglement
properties. Indeed, we have recently managed to solve the
separability problem for the case of tripartite system with one
mode per each party \cite{3party}. The result of Werner and Wolf
appeared first a little surprising in the view of Refs.
\cite{Duan00,Si00,Gi00}. Recently, some of us have been able to
clarify this and solve ultimately the separability \cite{sepg}
and distillability \cite{distg} problems for Gaussian states of
two parties sharing arbitrary number of modes. While the PPT
property remains a valid necessary and sufficient condition for
non-distillability, the separability criterion has a complex
form of a nonlinear map for covariance matrices.

Finally, it is worth mentioning that it is not known yet whether
there exist BES which do not have PPT, even though there is a
strong indication of this fact \cite{NPTBE}. If this were
finally true, this would have important implications in the
context of distillation \cite{SSTBE}, since it may well happen
that by mixing two NDS one obtains a distillable one.
%!
%%%%%%%%%%%%%%%%%%%%%%%%%%%%%%%%%%
\subsection{Question of genericity: the structural point of view}
%%%%%%%%%%%%%%%%%%%%%%%%%%%%%%%%%%
%!!
There is one open question whether the
%!
given
%!!
CV entanglement
represents a {\it generic} entanglement in the sense that it has
infinite Schmidt number (see \cite{Te00}), i. e. whether it is
the limit of matrices whose Schmidt number goes to infinity.
This means that, in principle, in order to generate the state,
one would have to be able to generate the states of arbitrary
Schmidt rank. However, some of the CV BES states similar to
``spurious'' ones could have also this property - if a finite
dimensional $n \otimes n $ PPT states with Schmidt rank of order
$O(n^{\alpha})$ with some $0<\alpha\le 1$ existed, then we could
put in the expression (\ref{oszuk}) the $k_{n} \otimes k_{n}$
states $\sigma_{n}$ with the rank $O(k^{\alpha}_{n})$ where say
$k_{n+1}=2(\sum_{i=1}^{n}k_{n})$. Thus, we see that in order to
describe the generic CV entangled states it seems reasonable to
require the stronger version the notion of infinite Schmidt
number. Intuitively, it should mean that the pure states with
infinite Schmidt rank are necessarily involved in the mixed
state representation. One possible definition would be that a
generic CV state with infinite Schmidt rank should be
necessarily of the form $\varrho=\sum_{i} p_{i} |\Psi_{i}
\rangle \langle \Psi_{i} | $, with $|\Psi_{i}\rangle$ not
necessarily orthogonal, but with
{\it at least one} $|\Psi_{i}\rangle$ of infinite rank. Such
states obviously exist -- take for instance one pure state of
infinte Schmidt rank, or a convex combination of two such
states. However, in the above definition the precise notion of
the decomposition in the CV case in the sense of Ref. \cite{Hu}
has to be specified. Another possible definition (which seems to
be significantly weaker) would be to require the generic CV
state to be the limit of $n \otimes n$ states of Schmidt rank
$n^{\alpha}$ for some $0<\alpha\le 1$.

Concerning BES -- we do not know whether there exists any BES
for CV with PPT property, having at the same time the feature of
being a generic CV state, whatever it would mean. It is worth
stressing at this point that according to the results of Ref.
\cite{Le00S}, PPT entangled state in $n \otimes n$ space are
expected to have Schmidt number smaller than $n$. In fact, in
the Appendix A, we present the arguments analogous to those used
in Ref. \cite{Le00S} that the typical PPT bound entangled states
in $n\otimes n$ space either have the Schmidt number of order
$O(1)$, or their partial transpose have this property. It is
possible, however, that the recently introduced Gaussian bound
entangled states \cite{Werner,sepg} satisfy all requirements as
far as the CV genericity is concerned. It would thus be
interesting to analyze the Schmidt number of those states.

%!
%%%%%%%%%%%%%%%%%%%%%%%%%%%%%%%%%%%%%%%%%%%%%%%%%%%%%%%%%%%%
\subsection{Question of genericity: distillation point of view}
%%%%%%%%%%%%%%%%%%%%%%%%%%%%%%%%%%%%%%%%%%%%%%%%%%%%%%%%%%%%

In former section we have dealed with question of genericity
of CV state as far as {\it the structure of the state
is concerned}. Bound (nondistillable)
entanglement is directly related to distillation procedures.
It is important to address the question from
different point of view i. e.
analysing the output  of distillation procedure
from the point of view of genericity.

The present result of section II clearly shows that
NDS in the sense of standard definition
(that no entanglment can be distilled
from given state) is nowhere dense
in set of CV states.
This is an important theorem generalising previous
results. However in the classical definition
of NDS treats both finite and infinite dimensional
entanglement nondistillable. For sake of many applications
the next step of study which would operationally distinguish
those two quantities would be desirable.

In such approach ``fully CV NDS '' would be all those
states that do not allow for distillation of
{\it infinite} pure entanglement (whatever it means).
This signifficantly increases the set
of states that are interpreted as bound entangled.

As we shall see below it is more complicated
issue. We will not give definite answers here.
However further methods of investigation will be suggested.

%The question is what would happen if we relaxed the
%definition of BES, or equivalently, put
%more restritions on notion of distillability to make the output
%of distillation process ``generically of CV kind''.
Again, as in previous section,
one of proposed definitions could be the following:

{\it A. The state $\varrho$ represents ``fully CV free (nondistillable)'' entanglement
if and only if it is possible (impossible)
to distill nonzero amount of pure states with infinite
Schmidt rank from state $\varrho$ }.

Nonzero amount is here understood in sense of
usual distillation yeld (i. e. as a nonzero amount of pairs).
Note that to qualify distilled entanglement in finite dimensions
the condition of asymptotic approaching
the maximally entangled state
$|\Psi_{+}\rangle=\frac{1}{\sqrt{m}} \sum_{i=1}^{m} |e_i ,e_i \rangle$
 was required.
It is known however that there is no maximally entangled states
of infinite Schmidt rank. Thus in place of $\Psi_{+}$ one would have probably
use some fixed pure state $\Psi_{\infty}$ having the reduced density matrix
nonsingular or at least of infinite rank (this is eqivalent
to the infinite Schmidt rank of $\Psi_{\infty}$).

Another interesting (weaker) defintion would be more in spirit of
Ref. \cite{Ho00} where increasing sequences of finite Schmidt
rank were used. Namely one can propose:

{\it B. The state $\varrho$ represents CV free (nondistillable) entangled
if and only if it is possible (impossible) to distill nonzero amount
$\eta_{p}>0$ of $p$-Schmidt rank states
with $limsup_{p}\eta_{p}>0$ .}

The main difficulty dealing with Schmidt rank in those definitions is
that the operational methods of its detection
in context of CV are not enough developed.

For example it is not known whether the proposal B above
is equivalent to the following generalisation
of the ``two-qubit subspace''(see sec. II A):

{\it the state is fully free (bound) entangled
iff there is (no) $n$ and the family of
bilocal filters $A_{p}\otimes B_{p}$ such that
the new $n$-copy states
\begin{equation}
 \varrho'_{p}=A_{p}\otimes B_{p}\varrho^{\otimes n} A_{p}^{\dagger} \otimes
 B_{p}
/ (A_{p}\otimes B_{p}\varrho^{\otimes n} A_{p}^{\dagger} \otimes B_{p})
\label{filtr}
\end{equation}
violate the $p$-Schmidt rank test via positive map
i. e. $[\id \otimes \Lambda_{p}](\varrho'_{p})$ is not positive matrix.}
The map $\Lambda_{p}(X)\equiv Tr(X) I - (p-1)^{-1}X$
is $p-1$-positive but not $p$-positive and
was used to detect $p$-Schmidt rank of isotropic entangled states
\cite{Te00}.

Dealing with the state (\ref{filtr}) is not easy because
even in the case of finite dimensions the possibility of
asymptotically singular denominator in formulas like (\ref{filtr})
leads to suprising effects (see \cite{single}). Putting
$A_{p} \otimes B_{p}$ equal identity we shall make the results of section II
slightly stronger and in the above spirit.

%%%%%%%%%%%%%%%%%%%%%%%%%%%%%%
\section{BES for CV are nowhere dense:
generalisation involving Schmidt numbers}
%%%%%%%%%%%%%%%%%%%%%%%%%%%%%

Here, using the trace norm topology from sec. II. B
we shall prove the stronger version of the main result
of sect. II. Suppose for moment that
as a ``fully CV free entanglement'' we shall treat {\it only}
very special CV states. Namely only those form which it
is possible to produce $p$-Schmidt rank state (with $p$ fixed
but {\it arbitrary} high) from {\it a single copy}
by means of special LOCC protocol given in
Appendix B. The protocol is a natural generalisation
of the protocol utilising reduciton criterion \cite{xor}.

Those special states form the set, say $D_{p}^{1}$
($1$ stands for ``single copy'' and $p$ for Schmidt rank).
This set is signifficantly smaller than the one
formed by the classical definition of free entangled states
(see sec. II. A.).
If as ``fully CV free states'' one treats the set
$D_{p}^{1}$ (which still seems to contain too much states,
c. f. discussion of sec. III.B, but this is for
dydactic purposes) than one enlarges the set
of what is understood as ``fully CV bound'' or ``fully CV nondistillable''.

Below we shall see that though the latter is larger it
is still nowhere dense. To have it we need to prove that
$D_{p}^{1}$ is (i) open and (ii) dense in set of all
density operators.
Any state $\varrho \in D^{1}_{p}$
satisfies by the very definition (see Appendix B)
the inequality:
\begin{equation}
\langle \Psi|[\id \otimes \Lambda_{p}](\varrho)| \Psi \rangle=\epsilon < 0.
\label{genreduction}
\end{equation}
Now suppose that $\varrho' \in
\tilde{B}_{\eta}(\varrho)=\{\varrho', ||\varrho - \varrho' ||_{T} < \eta \}$.
Then
\begin{eqnarray}
&&|\langle \Psi|[\id \otimes \Lambda_{p}](\varrho-\varrho')| \Psi \rangle|
=|\langle \Psi|(\varrho_{A} \otimes I - \varrho_{A}' \otimes I \nonumber \\
&&+ (p-1)^{-1}(\varrho-\varrho')|\Psi \rangle | \leq
 |Tr[\varrho_{\Psi}(\varrho_{A} - \varrho_{A}')]| + \nonumber \\
&&|\langle \Psi|(p-1)^{-1}(\varrho-\varrho')|\Psi \rangle |
 \leq Tr[\varrho_{A}^{\Psi}|\varrho_{A} - \varrho_{A}'|] + \nonumber \\
&& (p-1)^{-1} |\langle \Psi|\varrho-\varrho'|\Psi \rangle| \leq \eta (1+(p-1)^{-1})
\end{eqnarray}
where $\varrho_{A}^{\Psi}$ is reduced density
matrix of pure state $|\Psi\rangle \langle \Psi|$.

Where we have used two properties
(i) $||\varrho_{A}-\varrho_{A}'||_{T} \leq \eta$
because partial trace is tracepreserving
completely positive map which does not increase
the trace norm $||A||_{T}$;
(ii) $\langle \phi | A| \phi \rangle \leq ||A|| \leq ||A||_{T}$
for positive $A$ where $|| A ||$ stands for operator
norm as in sect. II.B.
Thus, for $\eta<\frac{|\epsilon|}{(1+(p-1)^{-1})}$,
any $\varrho' \in \tilde{B}_{\eta}(\varrho)$
satisfies $\langle \Psi|[\id \otimes \Lambda_{p}](\varrho')
| \Psi \rangle < 0$ i. e.  {\it ergo}
belongs to $D_{p}^{1}$. This
implies that the latter is open.
Now to prove that $D_{p}^{1}$ is dense one can repeat
the argument of sec. II.A with
$|\Psi_{N}\rangle \in  (H_{A} \ominus H_{A}^{N})
\otimes (H_{B} \ominus H_{B}^{N})$
being maximally entangled pure state of Schmidt rank $p$:
$|\Psi_{N}\rangle=\frac{1}{\sqrt{p}}\sum_{i=1}^{p}
|e_{i}, f_{i}\rangle$ (the only difference is that the resulting
state should be shown to satisfy (\ref{genreduction})
which is easy to see).
Thus $D_{p}^1$ is dense and open so its complement is nowhere
dense in set of all states which completes the proof.
Remarkable that the line of the proof remains completely correct for ``uniform''
assumption (i. e.  \ref{genreduction}) satisfied
with one $\epsilon$ for {\it all} natural $p$ some $|\Psi\rangle=|\Psi(p)\rangle$ )
but than the assumption itself can be easily shown to be {\it false}
in the sense that no state can satisfy it.
%!!

%%%%%%%%%%%%%%%%%%%%%
\section{Conclusions}
%%%%%%%%%%%%%%%%%%%%%

In this paper we have considered non--distillable states for
continuous variables. In the main part of the paper we have
proven that the subset of non--distillable states is nowhere
dense in the set of all CV states. This is a much stronger
result than the recent one by Clifton and Halvorson \cite{CH00},
which prove the same result for the set of separable states,
since that one is contained in the set of NDS. Moreover, our
results imply that the subsets of BES and PPT states are also
nowhere dense. Thus, generic CV states are distillable. We have
also presented some examples of BES and discussed their
genericity from the point of view of CV and their Schmidt
number. In the Appendix A we have presented an evidence that all
PPT BES in $n \otimes n$ systems either have Schmidt number
smaller that $O(1)$, or their partial transposes have this
property.
%!
Finally we have analysed the genericity
of CV entanglement in context of Schmidt number.
We studied the flexibility of the
main result under the assumptions
and proved more general result that
nowhere sende is the set of
all states from which it is impossible to
produce $p$-Schmidt rank state
from a single copy in some (well defined) way.
The latter involves single
copy protocol (provided in Appendix B)
being a generalisation of that obtained with help
of reduction cirterion. By provinding some
proposals of definitions of what can be treated as
``fully CV'' we have shown that further
investigation of genericity in context of CV in desirable.

We thank Anna Sanpera, Dagmar Bru\ss\, Geza Giedke and
Otfried G\"uhne for useful discussions. This work has been
supported by the DFG (SFB 407 and Schwerpunkt
``Quanteninformationsverarbeitung''), the Austrian Science
Foundation (SFB ``control and measurement of coherent quantum
systems''), the European Union Programme ``EQUIP''
(IST-1999-11053) and the Institute for Quantum Information GmbH.
Part of the work was completed at The Erwin Schr\"{o}dinger
International Institute for Mathematical Physics, during the
Program ``Quantum Measurement and Information'', Vienna 2000.

\appendix

\section{Schmidt number of PPT BES for $n \otimes n$ systems}

In this Appendix we essentially repeat the arguments used in the
Ref. \cite{Le00S} to support the conjecture that in $3\otimes 3$
systems all PPT BES have Schmidt number 2. We consider now the
$n \otimes n$ case, with $n$ large. Let $r(\rho)$ denotes the
rank fo $\rho$; our aim is to present a strong evidence for the
following conjecture:

{\it Conjecture .- All PPT entangled states in $n \otimes n$
systems either have Schmidt number of the order of $O(1)$ or
their partial transposes have this property.}

Note, that this conjecture concerns for instance projections of
the PPT BES (\ref{stan}) onto $n \otimes n$ spaces. We observe
that

\begin{itemize}

\item It is enough to show the conjecture for the, so called, edge
states \cite{substrac,iha1,iha2}, i.e. the PPT states $\delta$
such that there exist no product vector $|e,f\rangle$ in their
range, such that $|e^*,f\rangle$ is in the range of the
partially transposed operator $\delta^{T_A}$.

\item Let $r(\rho)$ denotes the rank of $\rho$. It is likely
that it is enough to prove the conjecture for the edge states of
maximal ranks \cite{substrac}, i.e. those whose ranks fulfill
$r(\delta) + r(\delta^{T_A})=2n^2-2n+1$. We expect that such
states are dense in the set of all edge states. To show the
latter statement, we consider an edge state $\tilde\delta$ which
does not have maximal ranks. We can always add to it
infinitesimal amount of projectors on product vectors destroying
the edge property. The resulting state $\rho$ would have more
product states in its range, than the product states used to
destroy the edge property. Subtracting projector on product
states different from the latter ones, would typically allow to
construct an edge state $\delta$ with maximal ranks, which would
be infinitesimally close to $\tilde\delta$ in any norm.

\item Let $R(A)$, $K(A)$  denotes the range and kernel of $A$,
respectively. The canonical form of an non--decomposable
entanglement witness that detects the edge state $\delta$ is
(\cite{iha1,iha2}, see also \cite{terh})
\begin{equation}
W=P+ Q^{T_A}-\epsilon \one,
\end{equation}
where the positive operators $P,Q$ have their ranges $R(P)=K(\delta)$,
$R(Q)=K(\delta^{T_A})$, and $\epsilon>0$ is sufficiently small so that
for any product vector $\langle e,f|W|e,f\rangle\ge 0$.

\item If we can show that for any edge state with maximal ranks
and any corresponding witness $W$ detecting its entanglement,
there exist a vector $|\psi^s\rangle$ of Schmidt number $s$ such
that $\langle \psi^s|W|\psi^s\rangle<0$, then we would conclude
that all edge states with maximal ranks, and thus all edge
states, and thus all PPT entangled state have the Schmidt number
$<s$. Equivalently, it is sufficient to show that $\langle
 \psi^s|W+\epsilon\one|\psi^s\rangle\le 0$.
\end{itemize}

Let us therefore try to construct the desired vector
$|\psi^s\rangle$ of Schmidt number $s$. In general such
(unnormalized) vector will have a form
\begin{equation}
 |\psi^s\rangle\propto\sum_{i=1}^s l_i|e_i,f_i\rangle,
\label{psis}
\end{equation}
where $l_i$ are arbitrary complex coefficients for
$i=1,\ldots,s$, and $|e_i,f_i\rangle$ are linearly independent
product vectors for $i=1,\ldots,s$. Note, that the vector
(\ref{psis}) depends on $s$ complex parameters $l_i$ for
$i=1,\ldots,s$, whereas each of the $s$ vectors $|e_i\rangle$,
$|f_i\rangle$ depends themselves of $n-1$ relevant complex
parameters.

Let $r(P)=k_1$, and $r(Q)=2n-1-k_1$. Since we want to prove the
conjecture either for the edge state $\delta$, or for its
partial transpose, without loosing the generality, we may assume
that $k_1\ge 1$. We may then single out one projector out of
$P$, and write $P=P_1+|\Psi\rangle\langle \Psi|$, where $P_1\ge
0$, $r(P_1)=r(P)-1$, and $|\Psi\rangle$ is in the range of $P$.
We can choose then $|e_i,f_i\rangle$ in such a way that
$Q|e_i^*,f_i\rangle=0$, and $P_1|e_1,f_i\rangle=0$. These are
effectively $2n-2$ equations for vectors $|e_i,f_i\rangle$ which
depend on $2n-2$ parameters, so that we expect a finite, but
quite large number of solutions (c.f. \cite{substrac}). At the
same time, $\langle\psi^s|Q|\psi^s\rangle$ will become a
quadratic hermitian form of $l_i$'s with vanishing diagonal
elements. Such a hermitian form has typically more than one
dimensional subspace $\cal N$ of negative eigenvalues for large
$s$. But, one has to fulfill also the last equation implied by
$\langle\psi^s |\Psi\rangle=0$; this limits the values of $l_i$
to a hyperplane, which should have at least one dimensional
common subspace with the subspace of negative eigenvalues $\cal
N$. This would prove that either the Schmidt number of $\delta$
or of $\delta^{T_A}$ is of the order of 1.

Note that for a given $\delta$, if the presented construction
can be shown to be successful for every witness of $\delta$,
then it provides a sufficient condition for the state $\delta$
to have the Schmidt number smaller than $s$.

%!
\section{Producing state
of $p$-Schmidt rank form single CV copy}

In this Appendix we briefly show how to produce
(by means of local operations and classical communication - LOCC)
the $p$-Schmidt rank state from any CV state violating
the separability condition
\begin{equation}
[\id \otimes \Lambda_{p}](\varrho)\geq 0.
\label{psep}
\end{equation}
Following Ref. \cite{Gi00}
this is further generalisation of the distillation
protocol of Ref. \cite{xor} where
the above criterion with  $p=2$ (called reduction criterion
c. f. \cite{Cerf}) has been used.
The separability tests of the form (\ref{psep})
(which can be called $p-1$-reduction criteria)
for $p>2$ considered first in \cite{Te00} are
examples of general positive maps separability tests (see \cite{sep}).

Consider some CV state $\varrho$ and
suppose that there exists $|\Psi\rangle$ such that
$\langle \Psi| [\id \otimes \Lambda_{p}](\varrho)|\Psi \rangle=\epsilon< 0 $.
Then following arguments of Ref. \cite{Gi00} we get that there must
exist $N$ such that for any $m>N$
the new $m \otimes m$ state $\varrho_{m}$
produced from  $\varrho$ by projection
onto the finitedimensional support of
${\cal H}_{m} \otimes {\cal H}_{m}$
satisfies
\begin{equation}
\langle \Psi| [\id \otimes \Lambda_{p}](\varrho_{m})|\Psi \rangle
\leq \frac{\epsilon}{2}< 0.
\label{in}
\end{equation}
Now instead of $|\Psi\rangle$ we put $|\Psi'\rangle$ being
a normalised projection of $|\Psi\rangle$ on support of
$\varrho_{m}$. In this way we get the inequality
identical to the one of \cite{xor} with the
only difference that $p$ was equal $2$ there.
This allows to repeat the reasoning of Ref. \cite{xor}:
after the application of suitable local filtering and
$U \otimes U^{*}$ twirling to $\varrho_{m}$
one produces the $m \otimes m$ isotropic state
$\varrho_{is}=(1-q) \frac{I}{m^{2}} + q |\Psi_{+}\rangle \langle \Psi_{+}|$
with the fidelity $F \equiv \langle \Psi_{+}|\varrho_{is}|\Psi_{+}\rangle>\frac{p-1}{m}$.
But the latter implies (see \cite{Te00}) that the final
state (produced from initial
$\varrho$ by means of local operations and classical communication - LOCC)
has  the Schmidt number at least $p$. This concludes the reasoning.
Note that further steps of recurrence protocol with
generalised XOR (\cite{xor}) can be applied.
%!!

\end{document}